\documentclass[manyauthors,nocleardouble,COMPASS]{cernphprep} 
\RequirePackage{lineno}
\usepackage{cernunits,cernchemsym,heppennames2,graphicx,subfig,amssymb,color,amsmath,bigstrut}
\usepackage{cite,fancyvrb}

\usepackage{lineno}
\newcommand{\dd}{\mathop{}\mathopen{}\mathrm{d}}

\usepackage{comment}
\usepackage{soul}

\begin{document}
\begin{titlepage}
\PHnumber{2022--292}

\PHdate{December 27, 2022}
\DEFCOL{CDS-Library}
\title{Transverse-spin-dependent azimuthal asymmetries of pion and kaon pairs produced in muon-proton and muon-deuteron semi-inclusive \\ deep inelastic scattering}
\Collaboration{The COMPASS Collaboration}
\ShortAuthor{The COMPASS Collaboration}

\begin{abstract}

A set of measurements of azimuthal asymmetries in the production of pairs of identified hadrons in deep-inelastic scattering of muons on transversely polarised ${}^{6}$LiD (deuteron) and NH$_{3}$ (proton) targets is presented. All available data  collected in the years 2002–2004 and 2007/2010 with the COMPASS spectrometer using a muon beam of $160\,\,\mbox{GeV}/c$ at the CERN SPS were analysed. 
The asymmetries provide access to the transversity distribution functions via a fragmentation function that in principle may be independently obtained from $e^+e^-$ annihilation data.
Results are presented, discussed and compared to existing measurements as well as to model predictions. Asymmetries of $\pi^{+}\pi^{-}$ pairs measured with the proton target as a function of the Bjorken scaling variable are sizeable in the range $x> 0.032$, indicating non-vanishing transversity distribution and di-hadron interference fragmentation functions. As already pointed out by several authors, the small asymmetries of $\pi^{+}\pi^{-}$ measured on the ${}^{6}$LiD target can be interpreted as indication for a cancellation of $u$ and $d$-quark transversity distributions.
\end{abstract}
\vfill
\Submitted{(to be submitted to Phys. Lett. B)}
\end{titlepage}

{\pagestyle{empty}
}

%
%

\section{Introduction}

The description of the nucleon spin structure remains one of the main challenges in hadron physics. For a polarised nucleon, a leading-twist description comprises eight transverse-momentum-dependent (TMD) parton distribution functions (PDFs), describing the distributions of longitudinal and transverse momenta of partons and their correlations with nucleon and quark polarisations~\cite{Anselmino:2020vlp}. After integration over quark intrinsic transverse momentum $k_{T}$, three PDFs fully describe the nucleon, \textit{i.e.} the momentum distribution function $f_1^q(x)$, the helicity distribution function $g_1^q(x)$ and the transversity distribution function $h_1^q(x)$, where $x$ denotes the Bjorken scaling variable. For simplicity, the latter will be referred to as "transversity" throughout this paper. While the momentum and the helicity distribution functions have been measured with good accuracy, the knowledge on transversity is inferior but steadily increasing. Unlike $f_1^q$ and $g_1^q$, transversity is not accessible at leading twist in inclusive deep-inelastic scattering (DIS) because it is related to soft processes correlating quarks with opposite chirality, making it a chiral-odd function. Transversity can be accessed through observables that conserve chirality, \textit{i.e.} when it is coupled to a chiral-odd partner. In this regard, measuring semi-inclusive deep-inelastic scattering (SIDIS) is advantageous as transversity is coupled to the chiral-odd fragmentation functions (FFs) that describe the hadronisation of a transversely polarised quark $q$ into unpolarised final-state hadrons. 

The major source of information on transversity has been measurements of transverse-spin-dependent asymmetries (TSAs) in single-hadron production in SIDIS ($\ell N^{\uparrow} \rightarrow \ell^{\prime}hX$),  where transversity is coupled to the Collins FF~\cite{Collins:1992kk}. Transverse-spin asymmetries define the size of the transverse-target-spin-dependent azimuthal modulations of the SIDIS cross section. 
The first measurement of the Collins asymmetries was performed by the HERMES Collaboration~\cite{Airapetian:2004tw} using a transversely polarised hydrogen target. Sizeable asymmetries were observed, suggesting non-zero transversity and Collins FFs. The COMPASS Collaboration has the highest statistics data in this field, \textit{e.g.} 28$\,$M pion pairs taken with the NH$_3$ (proton) target and 4$\,$M pion pairs taken with the $^{6}$LiD (deuteron) target. The COMPASS collaboration has delivered a full set of measurements, \textit{i.e.} both TSAs for unidentified charged hadrons~\cite{Alexakhin:2005iw} as well as pions and kaons~\cite{Alekseev:2008aa} using the transversely polarised deuteron target, and TSAs for unidentified charged hadrons~\cite{Alekseev:2010rw,Adolph:2012sn} as well as pions and kaons~\cite{Adolph:2014zba} using the transversely polarised proton target. 
The TSAs measured with the polarised proton target showed a non-zero signal for Collins asymmetries with high statistics and a wide kinematic coverage. The TSAs measured with the polarised deuteron target are compatible with zero indicating a possible cancellation between up and down quark contributions to transversity. Despite the lower accuracy of these data, they were shown to play a key role in the extraction of flavour-dependent transversity distribution functions and remain the only SIDIS measurement ever performed using a transversely polarised deuteron target. In order to complete the COMPASS programme~\cite{compass-ii}, a new high statistics measurement of TSAs using a transversely polarised deuteron target was performed in the last data taking campaign, in 2022.

A promising alternative approach to access transversity is the measurement of TSAs in semi-inclusive production of pairs of hadrons of opposite charge ($\ell\ N^{\uparrow} \rightarrow \ell^{\prime}h^{+}h^{-}X$). Following this approach, in this work $\pi^+\pi^-$ and $K^+K^-$ as well as $\pi^+K^-$ and $K^+\pi^-$ pairs will be studied. In this case, transversity is coupled to the chiral-odd interference fragmentation function~(IFF) $H_1^{\sphericalangle}$~\cite{Collins:1993kq,Jaffe:1997hf, Bianconi:1999cd}, which describes the hadronisation of a transversely polarised quark into a pair of unpolarised hadrons. At leading twist, and after integration over total transverse momentum, the differential cross section on a transversely polarised target comprises two terms and can be written as ~\cite{Bacchetta:2002ux}
\begin{equation}
\label{xsection}
\begin{split}
\frac{\dd^7\sigma}{\dd\cos\theta\dd M_{\rm{hh}}\dd\phi_{\rm{R}}\dd z\dd x\dd y\dd \phi_{\rm{S}}} = &\frac{\alpha^2}{2\pi Q^2y} \left( (1-y+\frac{y^2}{2})\sum_q e_q^2~f_1^q(x)~D_{1,q}(z,M_{\rm{hh}}^2,\cos\theta) \right. \\
&\quad \left. + S_{\perp}(1-y) \sum_q e_q^2\frac{|\textbf{p}_1-\textbf{p}_2|}{2M_{\rm{hh}}} \sin\theta~\sin\phi_{\rm{RS}}~h_1^q(x)~H_{1,q}^{\sphericalangle}(z,M_{\rm{hh}}^2,\cos\theta) \right).
\end{split}
\end{equation}

Here $\alpha$ is the fine-structure constant, $D_{1,q}(z,M_{\rm{hh}}^2,\cos\theta)$ is the spin-independent dihadron fragmentation function (DiFF), $y$ is the fraction of the lepton energy in the laboratory frame transferred to the exchanged virtual-photon and $Q^{2}$ the negative square of the four-momentum transfer. Here, $z$ is the fraction of the virtual-photon energy carried by the hadron pair, $M_{\rm{hh}}$ its invariant mass and $\theta$ the polar angle of the positive hadron with respect to the two-hadron boost axis in the two-hadron rest frame. The symbol $S_{\perp}$ denotes the component of the target spin vector $\bf{S}$ perpendicular to the virtual-photon direction, with $\phi_{\rm{S}}$ the azimuthal angle of the initial nucleon spin, $\phi_{S^{\prime}}$ the azimuthal angle of the spin vector of the fragmenting quark and $\phi_{\rm{RS}} = \phi_{\rm{R}} - \phi_{S^{\prime}} = \phi_{\rm{R}} + \phi_{\rm{S}} -\pi$. The azimuthal angle $\phi_{\rm{R}}$ is given as
\begin{equation}
\phi_{\rm{R}} = \frac{(\bf{q}\times\bf{l})\cdot \bf{R}}{|(\bf{q}\times\bf{l})\cdot\bf{R}|} \, \arccos \frac{(\bf{q}\times\bf{l})\cdot (\textbf{q}\times \bf{R})}{|\bf{q}\times\bf{l}|\, |\textbf{q}\times\bf{R}|},
\end{equation}
where  $\bf{l}$ is the incoming lepton momentum, \textbf{q} the virtual-photon momentum and \textbf{R} the relative hadron momentum defined as $\textbf{R} = (z_2\textbf{p}_1 - z_1\textbf{p}_2)/(z_1+z_2)$. 
The TSAs are experimentally accessible through the measured number of hadron pairs written as
\begin{equation}
N_{\rm{hh}}(x,y,z,M_{\rm{hh}}^2,\cos\theta,\phi_{\rm{RS}}) \propto \sigma_{\rm{UU}}(1 + f(x,y)P_{\rm{T}}D_{\rm{nn}}(y)A_{\rm{UT}}^{\phi_{\rm{RS}}}\sin\theta\sin\phi_{\rm{RS}}),
\end{equation}

where $f(x, y)$ is the target polarisation dilution factor, $P_{\rm{T}}$ is the transverse polarisation of the target nucleons and $D_{\rm{nn}}$ the transverse-spin transfer coefficient. A more detailed discussion about the theoretical framework can be found in Ref.~\cite{Adolph:2012nw}. As shown in Ref.~\cite{Adolph:2012nw}, the asymmetry
\begin{equation}
    A_{\rm UT}^{\sin\phi_{\rm RS}} = \frac{|\textbf{p}_1-\textbf{p}_2|}{2M_{\rm{hh}}} \frac{\sum_q\,e_q^2~h_1^q(x)~H_{1,q}^{\sphericalangle}(z,M_{\rm{hh}}^2,\cos\theta)}{\sum_q e_q^2~f_1^q(x)~D_{1,q}(z,M_{\rm{hh}}^2,\cos\theta)} 
\end{equation}
is proportional to the product of the transversity distribution function $h_1^q(x)$ and the polarised two-hadron interference fragmentation function $H_{1,q}^{\sphericalangle}(z,M_{\rm{hh}}^2,\cos\theta)$, summed over the quark flavours $q$ with charge $e_q$.

Transverse-spin-dependent asymmetries of hadron pairs were first measured by the HERMES Collaboration~\cite{Airapetian:2008sk} for pion pairs using a transversely polarised hydrogen target. A sizeable signal was seen as a function of $x$, indicating a sizeable $u$-quark transversity and non-vanishing interference fragmentation functions. 
The COMPASS collaboration has published measurements of transverse spin asymmetries for pairs of unidentified hadrons produced on polarised deuterons~\cite{Adolph:2012nw} and polarised protons~\cite{Adolph:2014fjw}. The COMPASS results obtained with the proton target showed significantly sizeable asymmetries and a clear slope in their $x$-dependence thanks to the high accuracy of the proton data set, while those  extracted from deuteron-target data were found to be compatible with zero. An intriguing similarity between Collins-like single-hadron asymmetries for the positive and negative hadrons extracted from the SIDIS hadron-pair data and the standard Collins asymmetries is observed as a function of \(x\), suggesting that both single hadron and hadron-pair transverse-spin dependent fragmentation functions are generated by the same elementary mechanism, as presented and discussed in Ref. ~\cite{Adolph:2015zwe}. 

In this paper, we present a new measurement of TSAs for identified hadron-pairs using the full data set collected by the COMPASS Collaboration on transversely polarised deuteron (2002-2004) and proton (2007 and 2010) targets.
The paper is organised as follows. Only a brief description of the experimental setup and data analysis are given in Sec.~\ref{sec:results}, as the same setup and methods of data cleaning, selection and extraction of TSAs as in previous COMPASS analyses~\cite{Adolph:2012nw,Adolph:2014fjw} are used. 
The measured asymmetries are presented in Sec.~\ref{sec:results} and discussed in Sec.~\ref{sec:discussion}. 

\section{Experimental data and analysis}
\label{sec:dataanalysis} 

The analysis presented in this paper is based on data collected in the years 2002-2004 and 2007/2010 using the COMPASS spectrometer~\cite{Abbon:2007pq} by scattering the naturally polarised $\mu^{+}$ beam of 160 GeV/$c$ delivered by the CERN SPS off transversely polarised ${}^{6}$LiD and NH$_{3}$ targets, respectively. For ${}^{6}$LiD, the average dilution factor calculated for semi-inclusive reactions is $\langle{f}\rangle\sim 0.38$ and the average polarisation is $\langle{P_{\rm T}}\rangle\sim 0.47$, while for NH$_{3}$ the corresponding values are $\langle{f}\rangle\sim 0.15$ and $\langle{P}_{\textrm{T}}\rangle\sim 0.83$, respectively.
The target consisted of two or three cylindrical cells assembled in a row, which can be independently polarised. 
In 2002–2004, two cells were used, each 60 cm long and 3 cm in diameter, separated by a 10 cm gap. 
In 2007 and 2010, the target consisted of three cells of 4 cm diameter, with gaps of 5 cm in between. The middle cell was 60 cm long and the two outer ones 30 cm long each. From 2006 on, a new solenoidal magnet was used to polarise the target with a polar angle acceptance of 180 mrad as seen from the upstream end of the target, while in the earlier measurements with the ${}^6$LiD target the polar angle acceptance was 70 mrad. 
For the measurement of transverse spin effects, the target material was polarised along the vertical direction. In order to reduce systematic effects, neighbouring cells were polarised in opposite directions allowing for simultaneous data taking with both target spin directions to reduce flux-dependent systematic uncertainties.
Furthermore, the polarisation was destroyed and built up in reversed direction every four to five days, in order to cancel residual acceptance effects associated with
the longitudinal position of the target cells (\textit{i.e.} position along the beam line).
For the data collected using a proton target, in the analysis, the central cell is divided into two parts, providing four data samples with two different orientations of polarisation. Note that for the measurements in 2007 and in 2010 a similar spectrometer configuration was used. 

In the analysis, events with incoming and outgoing muons and at least two reconstructed charged hadrons originating from the interaction vertex inside the target cells are selected. Equal flux through the whole target is obtained by requiring that the extrapolated beam tracks pass through all three cells. In order to select events in the DIS regime, requirements are applied on the squared four-momentum transfer, $Q^{2}>1$ (GeV/$c)^{2}$, and on the invariant mass of the final hadronic state, $W>5~\mbox{GeV}/c^{2}$. Furthermore, the fractional energy transfer to the virtual photon is required to be $y>0.1$ and $y<0.9$ to remove events with poorly reconstructed virtual-photon energy and events with large radiative corrections, respectively.

For a selected DIS event, all reconstructed hadrons originating from the interaction vertex are considered. Only hadrons produced in the current fragmentation region are selected by requiring $z > 0.1$ for the fractional energy and $x_F > 0.1$ for the Feynmann-$x$ variable. The two-hadron sample consists of all combinations of oppositely charged hadrons built from the same DIS event. 
Exclusive dihadron production is suppressed by requiring the missing energy for each hadron pair to be greater than 3 GeV. As the azimuthal angle $\phi_{\rm{R}}$ is only defined for non-collinear vectors $\textbf{R}$ and $\textbf{q}$, a minimum value is required on the component of $\textbf{R}$ perpendicular to $\textbf{q}$, \textit{i.e.} $R_{\perp} > 0.07$ GeV/$c$. After the application of all requirements, 0.56 $\times\, 10^{7}$ h$^{+}$h$^{-}$ combinations remain for the deuteron data and 3.5 $\times\,10^{7}$ h$^{+}$h$^{-}$ pairs for the proton data.

\begin{table}[b]
    \centering
    \caption{Final statistics for unidentified and identified charged-hadron pairs in deuteron (2002-2004) and proton (2007 and 2010) data.}
    \begin{tabular}{lcccccc}
        \hline
        Year & \multicolumn{5}{c}{Number of pairs ($\times 10^{6}$)} \bigstrut[tb]\\ \cline{2-6}
        & h$^{+}$h$^{-}$ & $\pi^{+}\pi^{-}$ & $\pi^{+}K^{-}$ & $K^{+}\pi^{-}$ & $K^{+}K^{-}$ \bigstrut[tb]\\ \hline
        2002-2004 (deuteron) & 5.65 & 3.97 & 0.26 & 0.30 & 0.10 \\ 
        2007 (proton) & 10.91 & 7.41 & 0.38 & 0.53 & 0.22 \\
        2010 (proton) & 34.56 & 20.60 & 1.10 & 1.53 & 0.60 \\ \hline
    \end{tabular}
    \label{tab:final_stat}
\end{table}
\begin{figure}[htp]
  \centering
  \includegraphics[trim=0 0 0 0,clip,width=0.31\textwidth]{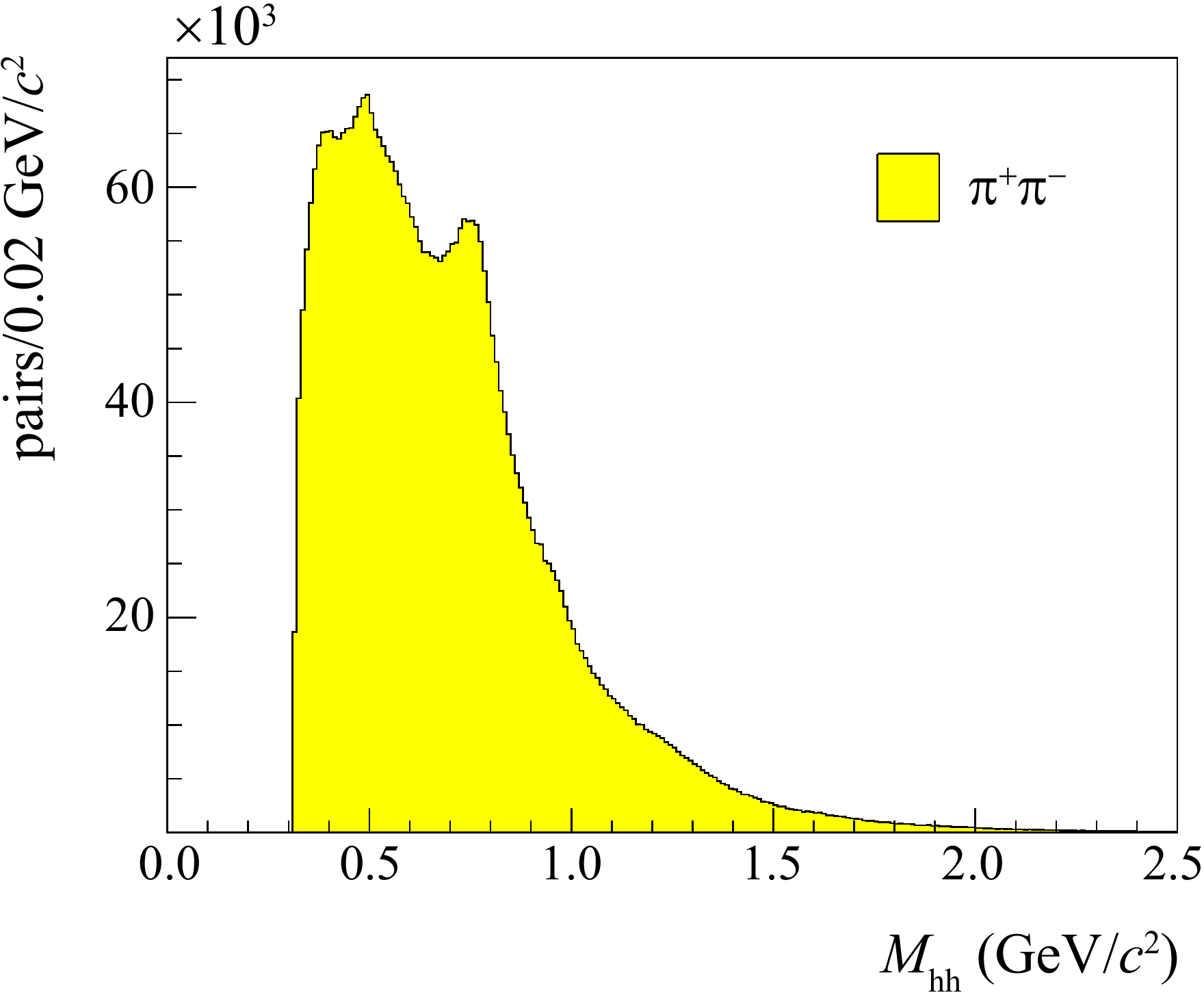}
  \includegraphics[trim=0 0 0 0,clip,width=0.31\textwidth]{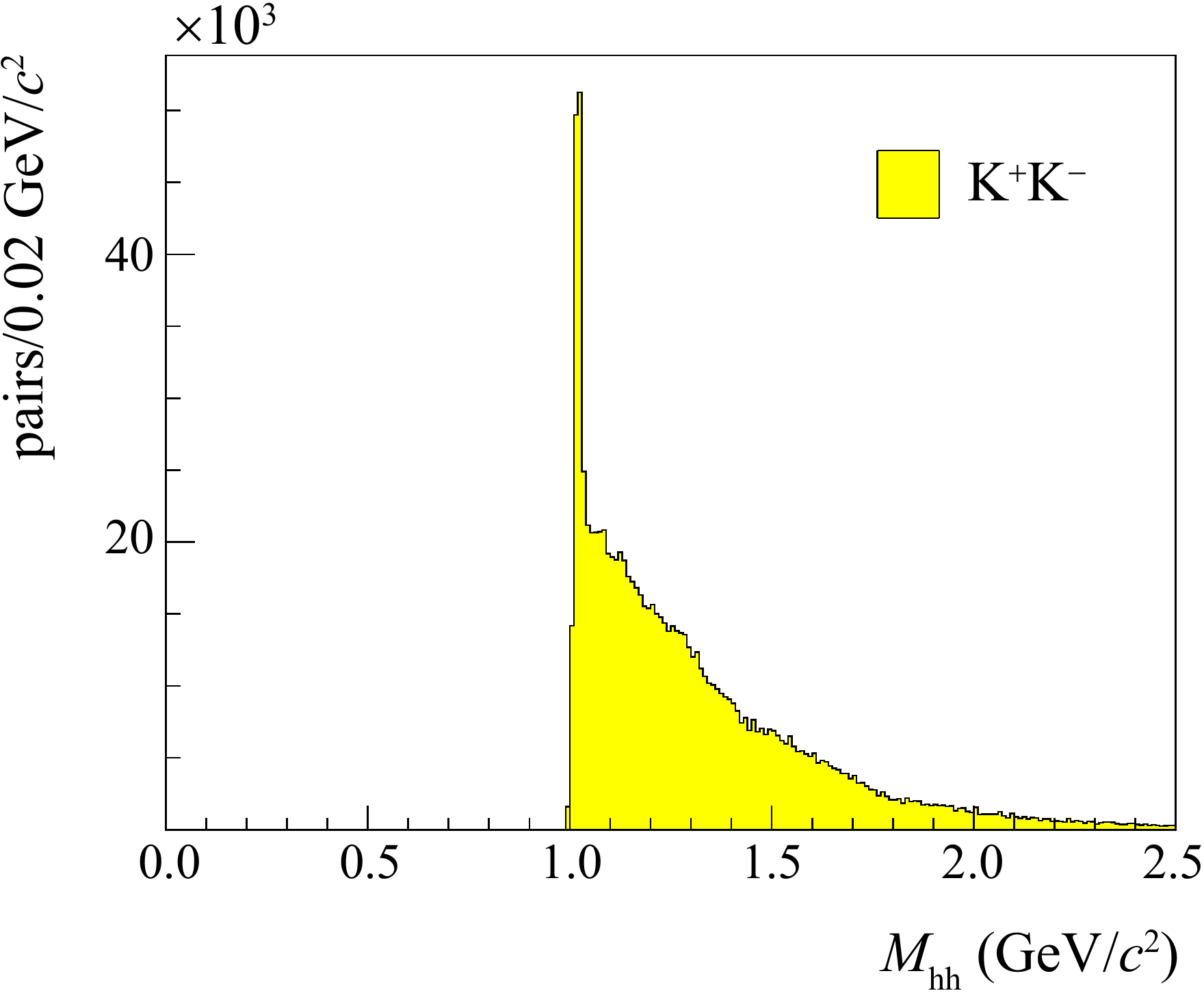}
  \includegraphics[trim=0 0 0 0,clip,width=0.31\textwidth]{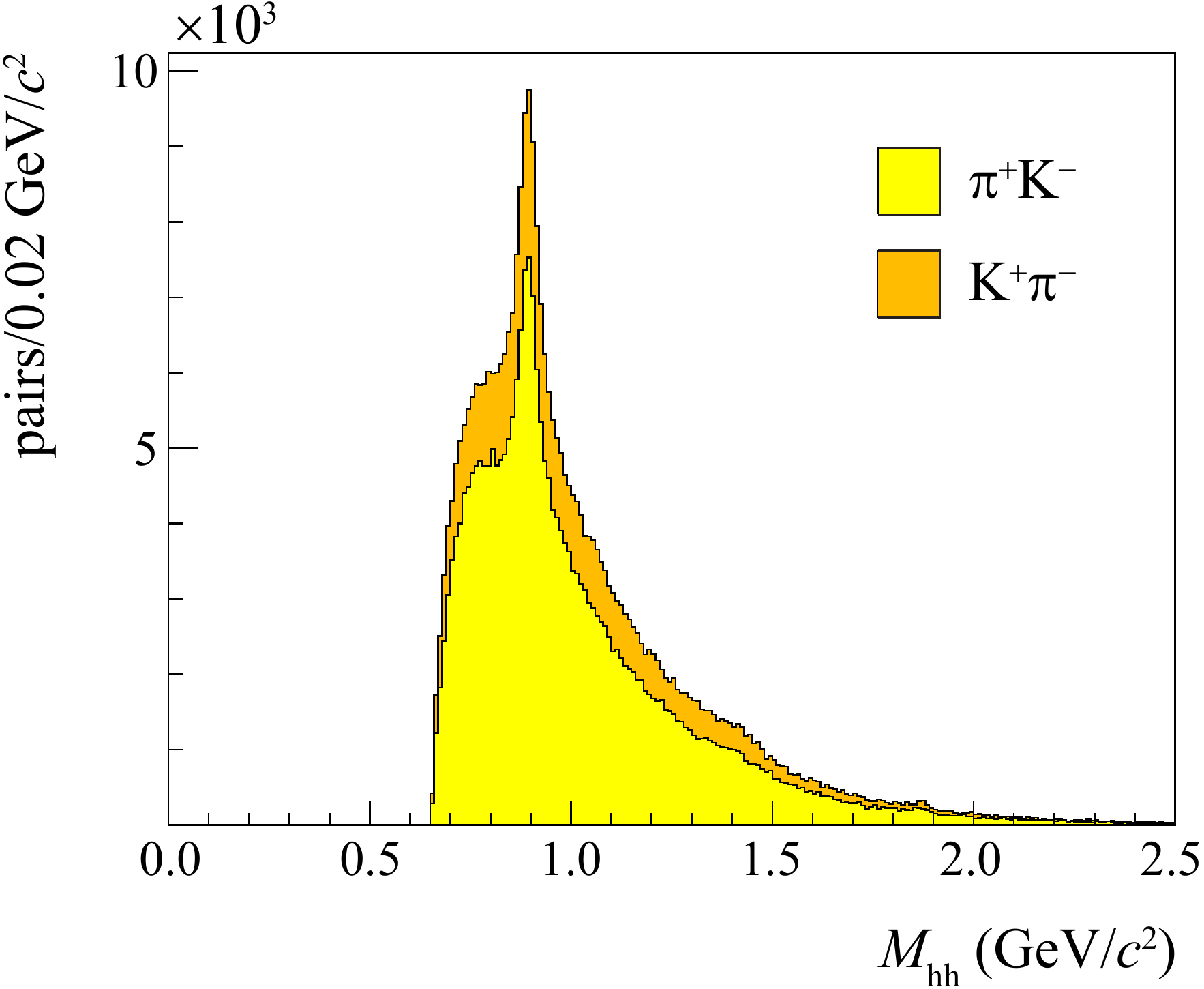}
  \includegraphics[trim=0 0 0 0,clip,width=0.31\textwidth]{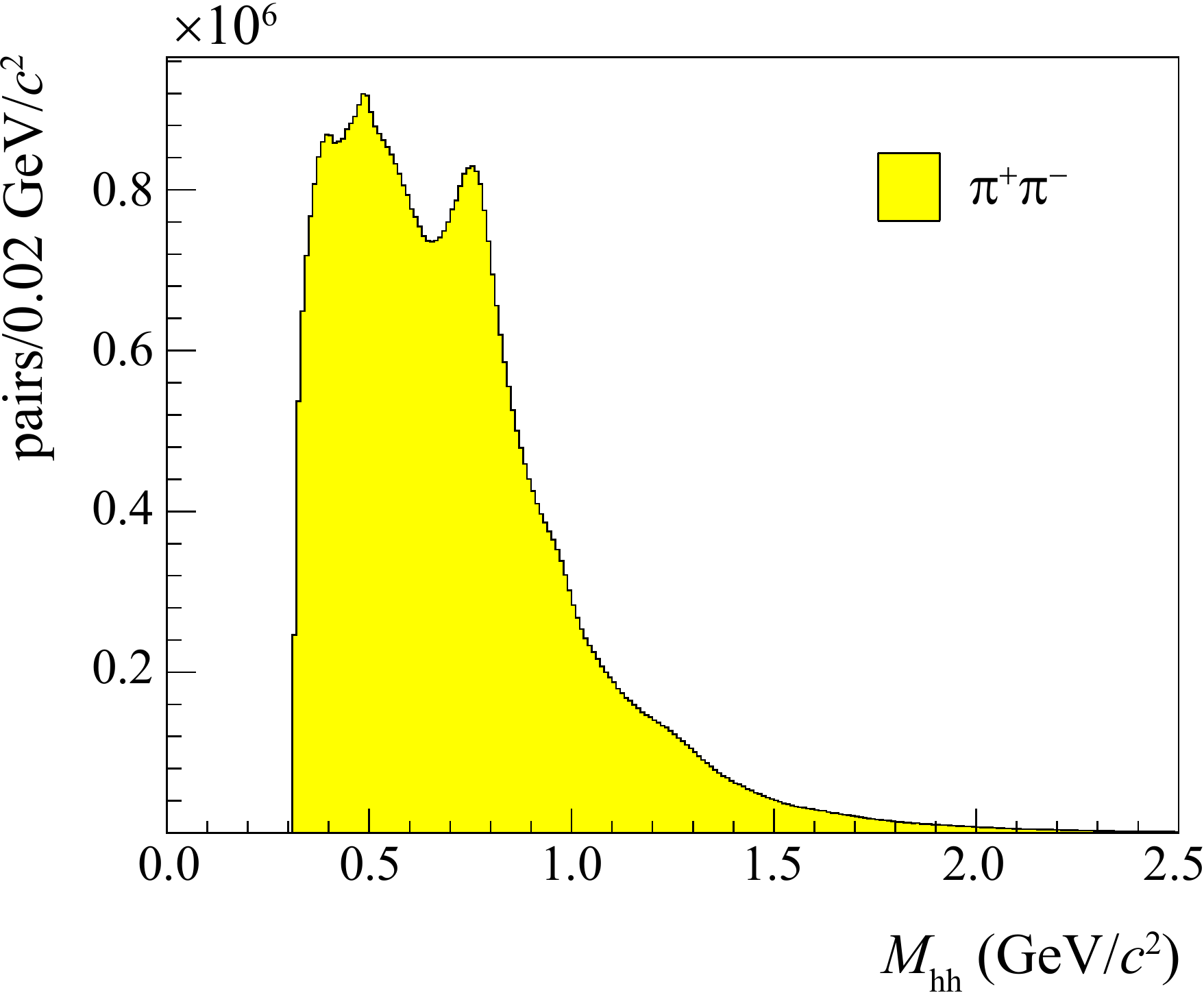}
  \includegraphics[trim=0 0 0 0,clip,width=0.31\textwidth]{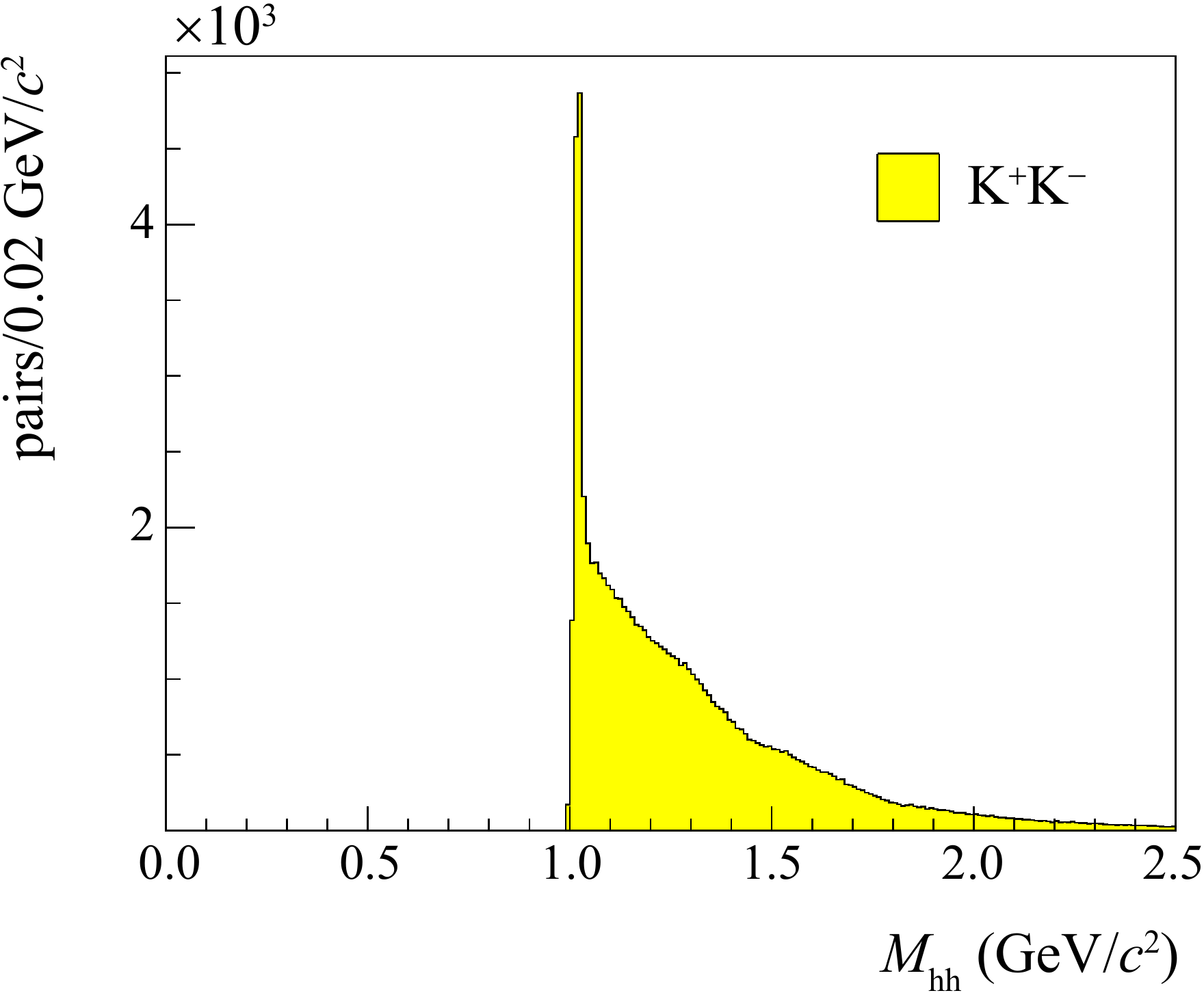}
  \includegraphics[trim=0 0 0 0,clip,width=0.31\textwidth]{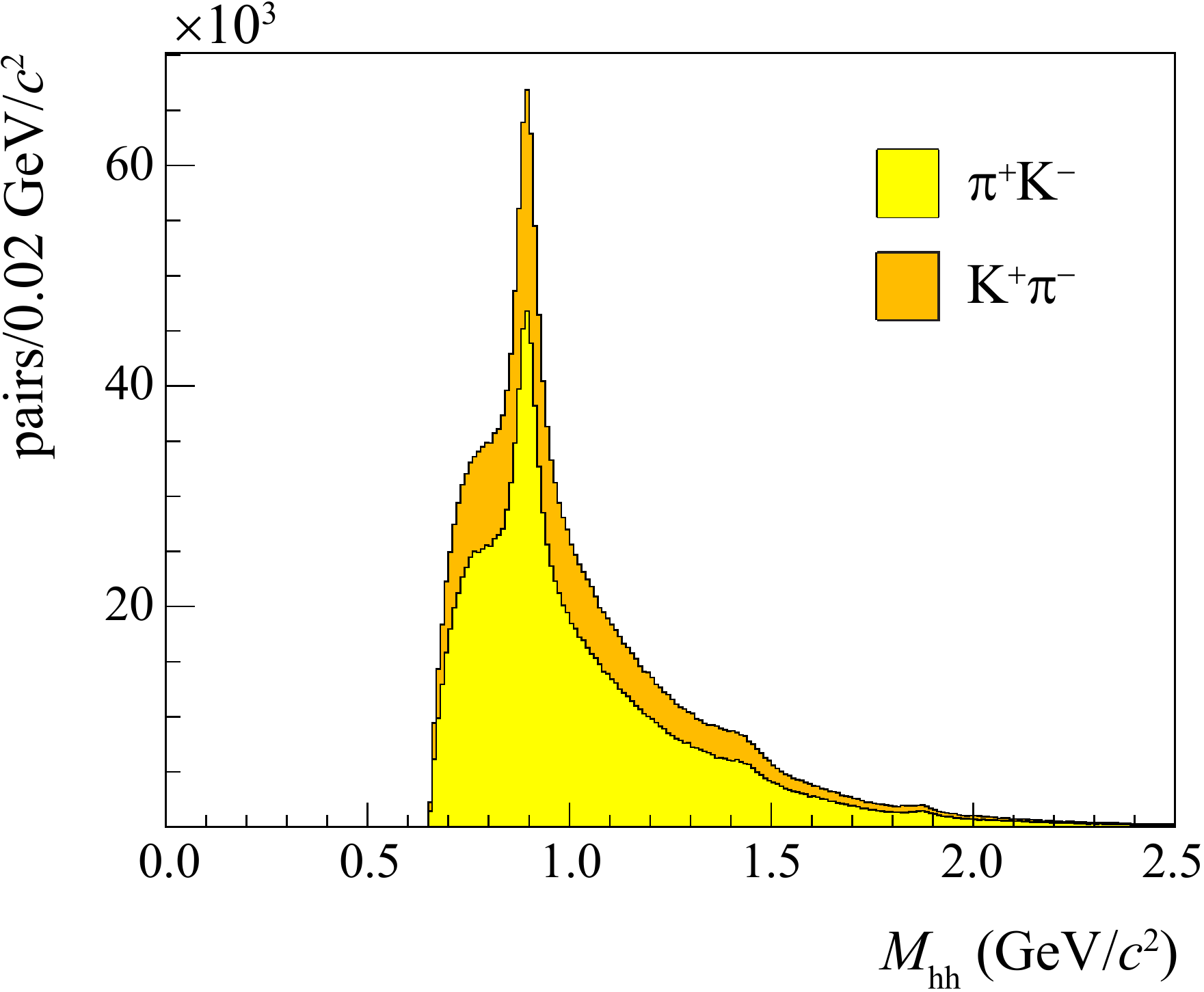}
  \caption[Deuteron and proton $M_{\rm{hh}}$ distributions]{Distributions of invariant mass $M_{\rm{hh}}$ for $2002$-$2004$ 
  deuteron data (top row) and combined $2007$/$2010$ 
  proton data (bottom row): $\pi^{+}\pi^{-}$ pairs ($1^{\text{st}}$ column)
  , $K^{+}K^{-}$ pairs ($2^{\text{nd}}$ column), $\pi^{+}K^{-}$ and $K^{+}\pi^{-}$ pairs ($3^{\text{rd}}$ column).\label{fig:mass}
  }
\end{figure}

The RICH detector information is used to identify charged hadrons as pions or kaons in the momentum range between the Cherenkov threshold (about 2.6 GeV/$c$ and 9 GeV/$c$, respectively) and 50 GeV/$c$. The detector set-up after the upgrade of 2005 and the particle identification (PID) procedure are fully described in Ref.~\cite{ABBON201126}, while details on the likelihood PID method and the purity of identified samples are explained in Ref.~\cite{Alekseev:2008aa} and Ref.~\cite{Adolph:2014zba} for deuteron and proton targets, respectively.
In the kinematic domain of the COMPASS experiment, about $67\,\%$ of the final-state charged hadrons are identified as pions and about $10\,\%$ as kaons. The remaining particles are either protons, electrons or not clearly identified. About $60\,\%$ are pion pairs ($\pi^{+}\pi^{-}$), about $2\,\%$ are kaon pairs ($K^{+}K^{-}$) and about $8\,\%$ are mixed pairs ($\pi^{+}K^{-},~K^{+}\pi^{-}$). The missing fraction refers to cases where at least one of the two hadrons cannot be accurately identified. The resulting statistics for unidentified and identified hadron pairs after applying all requirements are shown in Table~\ref{tab:final_stat}.

The invariant-mass distributions for the four opposite-charge combinations that can be formed using identified charged pions and kaons ($\pi^{+}\pi^{-}$, $K^{+}K^{-}$, $\pi^{+}K^{-}$, $K^{+}\pi^{-}$) are shown in Fig.\ref{fig:mass} for deuteron and proton targets. In the $\pi^{+}\pi^{-}$ spectrum, the mass signatures of some mesons decays, such as $K^{0}$ around $500\,\mbox{MeV}/c^{2}$, $\rho^0$ around $770\,\mbox{MeV}/c^{2}$, $f_{0}$ around $980\,\mbox{MeV}/c^{2}$ and $f_{2}$ around $1270\,\mbox{MeV}/c^{2}$, respectively, are clearly visible in both deuteron and proton data as expected from Ref.~\cite{Bacchetta:2006un}. Other decays with more than two hadrons in the final state (like the decays of $\omega$, $\eta$ and $\eta^{\prime}$) generate broader peaks and contribute less to the overall pion-pair invariant-mass spectra~\cite{Bacchetta:2006un}.  
The $K^{+}K^{-}$ invariant-mass distribution shows a very pronounced signal of the $\phi(1020)$ resonance close to its production threshold.  The $\phi$ meson can also contribute to the pion pair spectra via the two-step decay $\phi(1020) \rightarrow \rho \pi \rightarrow \pi^{+} \pi^{-} \pi^{0}$. The invariant-mass distribution of $K^+K^-$ pairs in the proton data shows indications of further broad peaks around $1300\,\mbox{MeV}/c^{2}$ and $1500\,\mbox{MeV}/c^{2}$, which might be caused by $f_{2}(1270)$ and $f_{2}^{\prime}(1525)$. The invariant-mass distributions of $\pi^{+}K^{-}$ and $K^{+}\pi^{-}$ also show in each case one dominant channel caused by the decays of $K^{*}(892)$.  Further possible candidates for peaks in the $M_{\rm{hh}}$ spectra of the $\pi^+K^-$ and $K^+\pi^-$ pairs are $K^{*}(1430)$ and $K_{4}^{*}(2045)$.
%


\section{Results}
\label{sec:results} 

\begin{figure}[t]
  \begin{center}
  \includegraphics[width=0.85\textwidth,trim= 0 0 0 0] {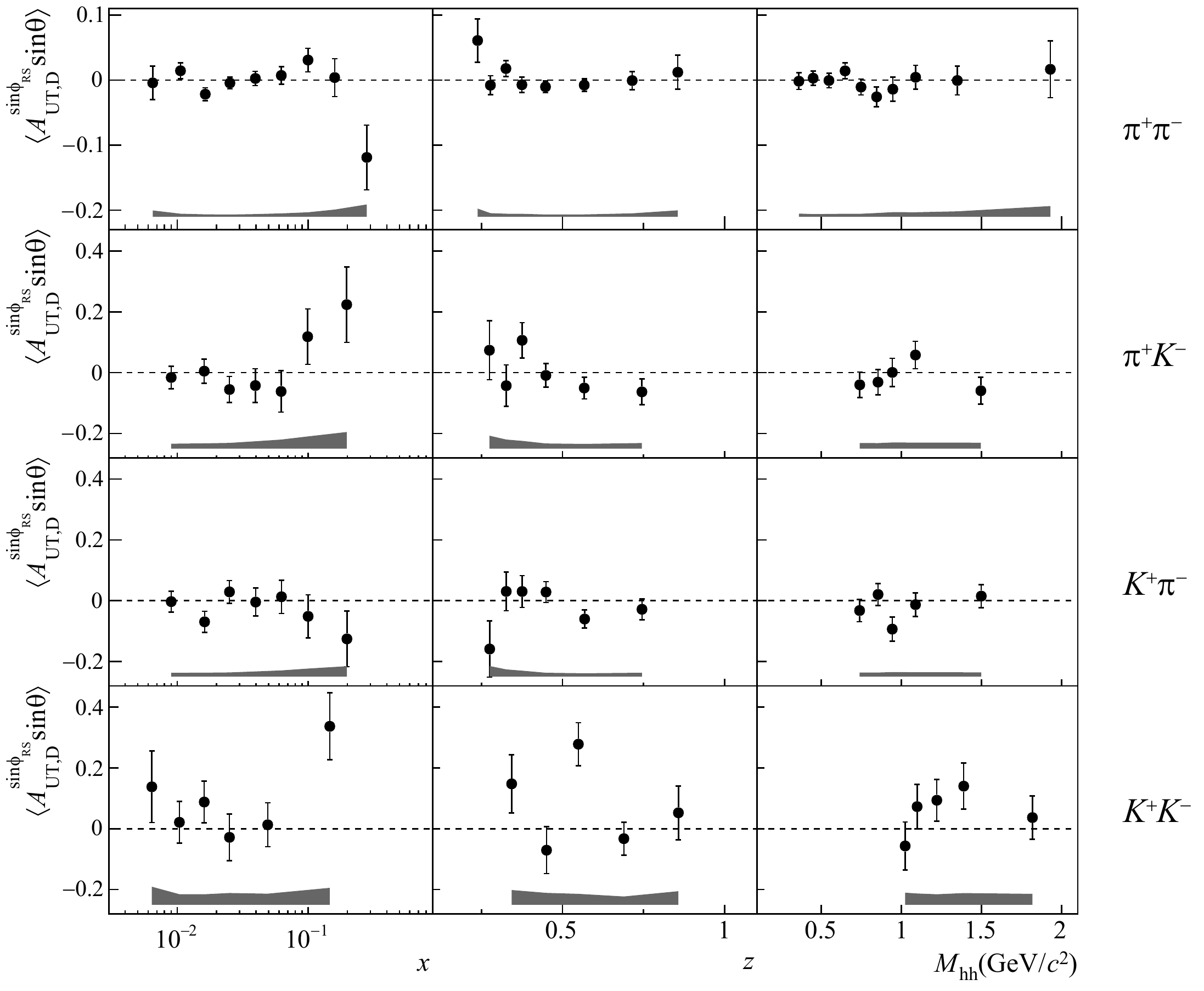}
  \caption{Hadron-pair transverse-spin-dependent asymmetries as a function of $x$, $z$ and $M_{\rm{hh}}$, extracted from the full data set collected with the ${}^{6}$LiD (deuteron) target. Systematic uncertainties are shown by the gray bands.}
  \label{pic:final_asyms_sys_d}
  \end{center}
\end{figure}
\begin{figure}[!t]
  \begin{center}
  \includegraphics[width=0.85\textwidth,trim= 0 0 0 0] {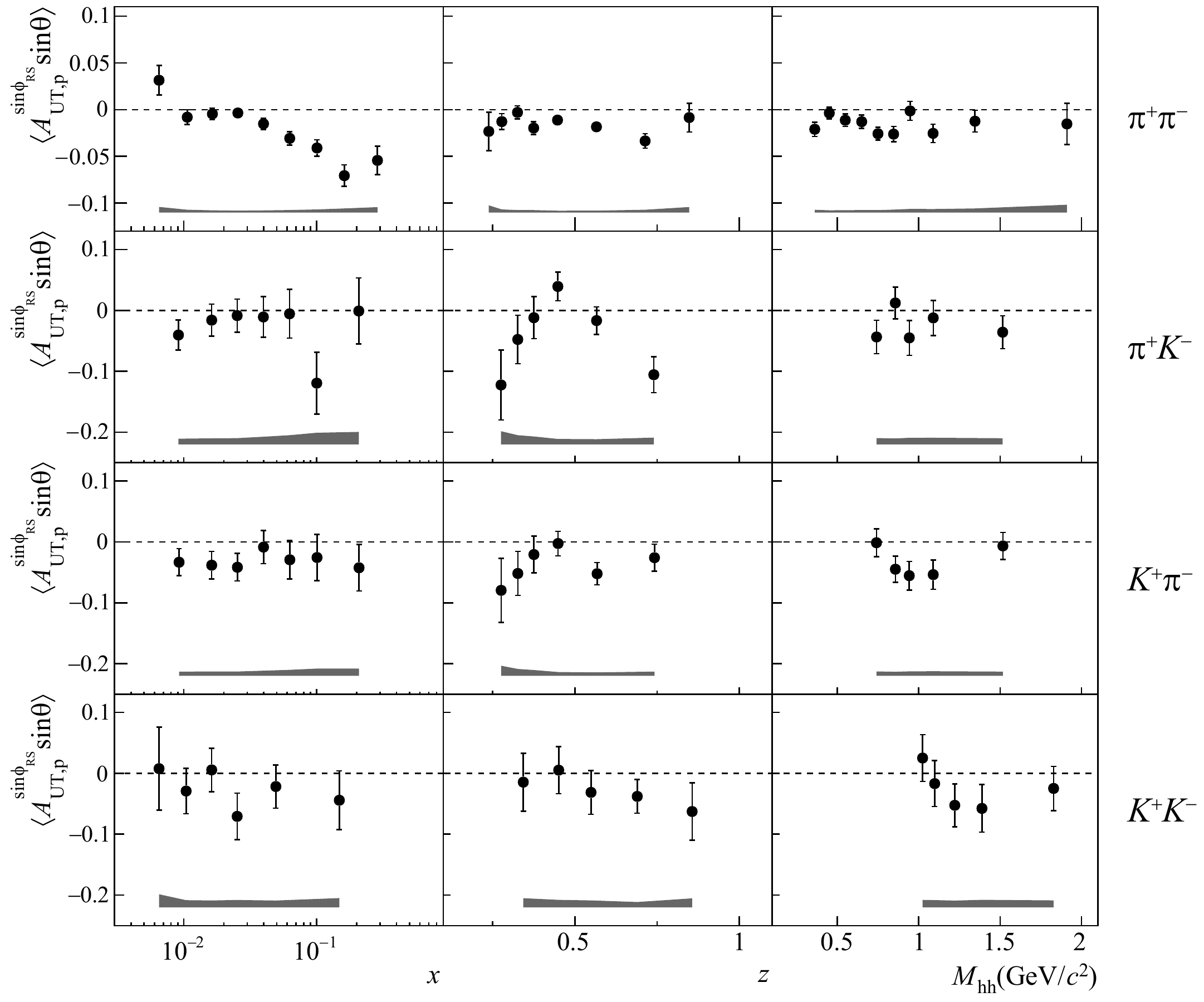}
  \caption{Hadron-pair transverse-spin-dependent asymmetries as a function of $x$, $z$ and $M_{\rm{hh}}$, extracted from the full data set collected with the NH$_{3}$ (proton) target. Systematic uncertainties are shown by the gray bands.}
  \label{pic:final_asyms_sys_p}
  \end{center}
\end{figure}

The asymmetries extracted from $^6$LiD and NH$_3$ targets are presented in Figs.~\ref{pic:final_asyms_sys_d} and~\ref{pic:final_asyms_sys_p}, respectively. They were evaluated in bins of $x$, $z$ and $M_{\rm{hh}}$ as given in Table~\ref{tab:binning}. 
For $^6$LiD, no significant asymmetry is observed in any variable for all pair combinations. For $\mathrm{NH_3}$, large negative asymmetries up to $-0.07$ are obtained for $\pi^+\pi^-$ pairs in the region $x > 0.03$, which implies that both transversity distributions and polarised two-hadron interference fragmentation functions do not vanish, as already observed in Refs.~ \cite{Airapetian:2008sk,Adolph:2014fjw}. For $x < 0.03$, these asymmetries are compatible with zero. 
\begin{table}[b]
    \centering 
    \caption{Bin limits of the variables $x$, $z$ and $M_{\rm{hh}}$ (in units of $\,\mbox{GeV}/c^{2}$) for the four types of pairs.}
    \begin{tabular}{c|ccccccccccc} 
        \hline
        & \multicolumn{11}{c}{$x$ bin limits} \bigstrut[tb]\\ \cline{2-12}
        $\pi\pi$ & 0.003 & 0.008 & 0.013 & 0.020 & 0.032 & 0.050 & 0.080 & 0.130 & 0.210 & 1.000 \\
        $\pi K/K\pi$ & 0.003 & 0.013 & 0.020 & 0.032 & 0.050 & 0.080 & 0.130 & 1.000 \\
        $K K$ & 0.003 & 0.008 & 0.013 & 0.020 & 0.032 & 0.080 & 1.000 \\ 
        \hline
        & \multicolumn{11}{c}{$z$ bin limits} \bigstrut[tb]\\ \cline{2-12}
        $\pi\pi$ & 0.20 & 0.25 & 0.30 & 0.35 & 0.40 & 0.50 & 0.65 & 0.80 & 1.00 & \\
        $\pi K/K\pi$ & 0.20 & 0.30 & 0.35 & 0.40 & 0.50 & 0.65 & 1.00 \\
        $K K$ & 0.20 & 0.40 & 0.50 & 0.65 & 0.80 & 1.00 \\ 
        \hline  
        & \multicolumn{11}{c}{$M_{\rm{hh}}$ bin limits} \bigstrut[tb]\\ \cline{2-12}
        $\pi\pi$ & 0.0 & 0.4 & 0.5 & 0.6 & 0.7 & 0.8 & 0.9 & 1.0 & 1.2 & 1.6 & 100 \\
        $\pi K/K\pi$ & 0.0 & 0.8 & 0.9 & 1.0 & 1.2 & 100 \\
        $K K$ & 0.9 & 1.05 & 1.15 & 1.30 & 1.50 & 100 \\
        \hline
    \end{tabular}
    \label{tab:binning}
\end{table}
%
%
The asymmetry measured with the $^6$LiD target is compatible with zero within uncertainties over the whole $x$ range. For both targets, no clear dependence on $z$ can be observed, and for the NH$_3$ target the asymmetry is observed to be negative in the whole range. For both targets, the $M_{\rm hh}$-dependence shows negative asymmetry values in the region of the $\rho^0$ mass.


For $K^{+}K^{-}$ pairs, the proton data show negative asymmetries in all three variables, while the deuteron data show indications for a positive signal. In particular the $M_{\rm{hh}}$-dependence shows opposite signs for the asymmetries measured with the NH$_{3}$ and ${}^{6}$LiD target, with an indication of a mirror-symmetric shape.
In the case of $\pi^{+}K^{-}$ and $K^{+}\pi^{-}$ pairs, the deuteron data show asymmetries compatible with zero, while the proton data show slightly negative asymmetries. 



\begin{figure}[tb]
 \centering 
 \includegraphics[width=0.8\textwidth, clip=true, trim = 0 0 0 0]{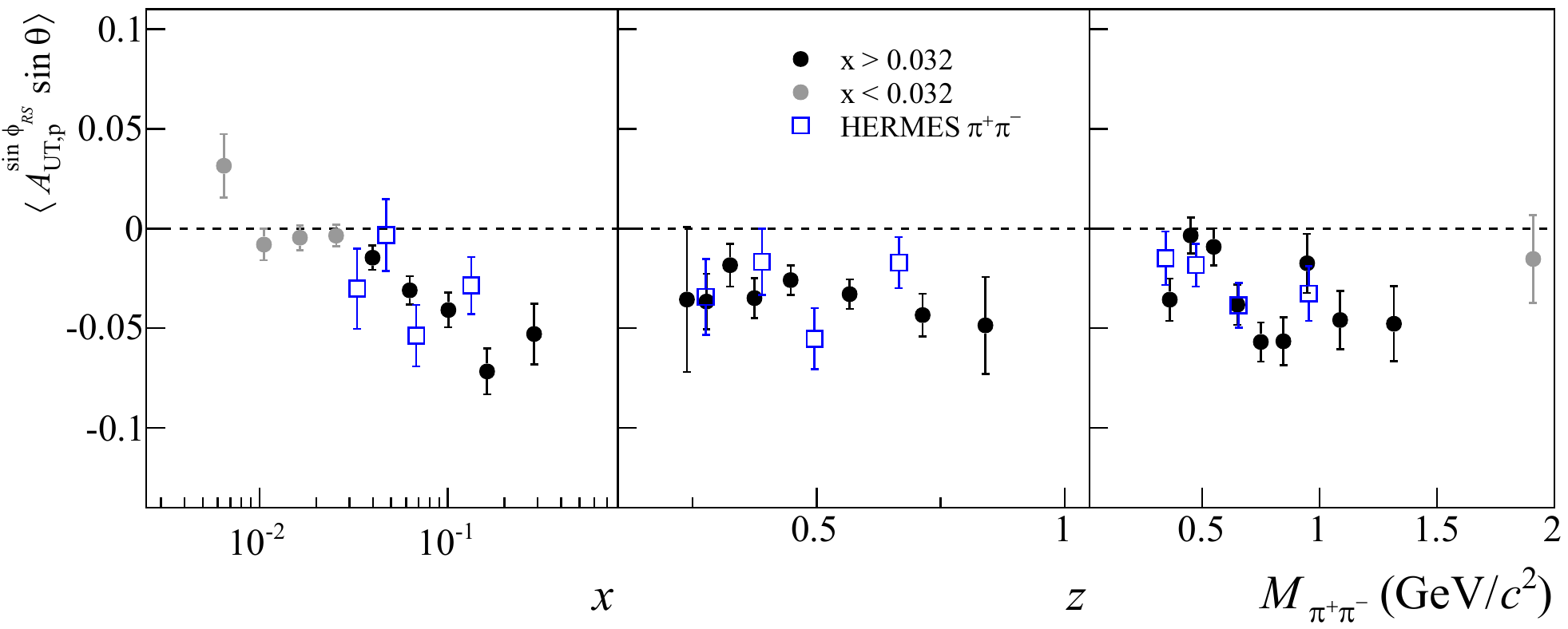}
 \caption[Asymmetry of $\pi^{+}\pi^{-}$ pairs: HERMES vs. COMPASS]{Comparison of $\pi^{+}\pi^{-}$ pair asymmetries measured by the HERMES Collaboration~\cite{Airapetian:2008sk} (blue open squares) with the results of the COMPASS Collaboration re-evaluated in the $x> 0.032$ region (black dots). \label{fig:dihadron_COMPASS_proton0710_HERMES}}
\end{figure}

The HERMES Collaboration measured TSAs for $\pi^{+}\pi^{-}$ pairs using electron-proton scattering~\cite{Airapetian:2008sk}. Given the wider kinematic coverage by COMPASS, the $\pi^+\pi^-$ COMPASS asymmetry was re-evaluated in the region $x > 0.032$ to allow for a direct comparison. The comparison is shown in Fig.~\ref{fig:dihadron_COMPASS_proton0710_HERMES}.
The results are in very good agreement within statistical uncertainties. 

\section{Interpretation of the results}
\label{sec:discussion}

The dihadron fragmentation functions entering the SIDIS cross section in Eq. (1) are non-perturbative objects. As such, they can not be calculated from first principles. Two classes of models have been proposed to describe them. In spectator-jet type models a mechanism different from that of the Collins FF is invoked to produce a non-vanishing $H_1^{\sphericalangle}$ function. Such a mechanism involves the interference between the amplitudes of two competing channels for the production of the hadron pair, \textit{e.g.} either the amplitude for direct production and the amplitude for resonance production \cite{Collins:1994ax,Bacchetta:2006un}, or the two amplitudes for the production of two different resonances \cite{Jaffe:1997hf}.
A different approach is followed by the recursive string+${}^3P_0$ model of polarised quark fragmentation \cite{kerbizi-2019}. It is implemented in the StringSpinner package \cite{Kerbizi:2021pzn} for the simulation of the Collins effect for pseudoscalar mesons produced in the fragmentation of transversely polarised quarks in SIDIS with the PYTHIA 8 event generator \cite{pythia8}.


\begin{figure}[b]
\vspace{-1em}
\centering
\includegraphics[width=0.7\textwidth]{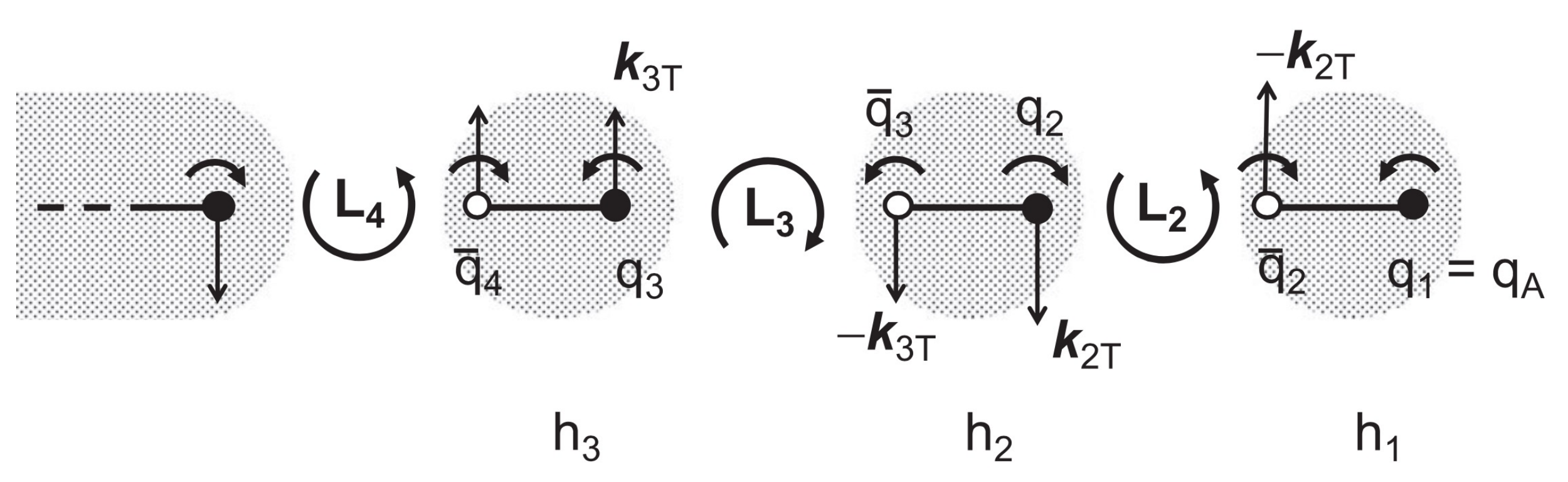}
\caption{The string+${}^3P_0$ mechanism of polarised quark fragmentation \cite{Kerbizi:2018qpp}. The closed (open) circles represent quark (antiquarks) at the string ends. The circular arrows above quarks show the orientation of their spins whereas the arrows at each string breaking $L_2, L_3\dots$ represent the orientation of the relative orbital angular momenta of the $q\bar{q}$ pairs. The straight arrows indicate the quark transverse momenta.}\label{fig:string+3p0}
\end{figure}

The classical string+${}^3P_0$ model for the fragmentation of a transversely polarised quark $q_A$ is illustrated in Fig. \ref{fig:string+3p0}. The string is stretched between the scattered quark $q_A$ and the target remnants along the quark direction and the string fragmentation occurs via tunneling of quark-antiquark pairs in the ${}^3P_0$ state, \textit{\textit{i.e.}} with spin $S=1$ and relative orbital angular momentum $L=1$, such that the total angular momentum $J$ is zero. Given the polarisation of $q_A$, taken here along the normal to the figure plane, at the string breakings the spin and the transverse momentum of the quark and antiquark, as well as the transverse momentum of the produced hadron are fixed. The rank $r$ indicates how far the hadron $h_r$ is produced from the fragmenting quark $q_A$, with $h_1$ being the hadron which contains $q_A$. For odd (even) $r$ the hadron $h_r$ is emitted to the left (right) with respect to the plane spanned by the momentum and polarisation vectors of the fragmenting quark. As an example, if the flavor of the fragmenting quark is $q_A=u$ and $h_1=\pi^+$, it can be $h_2=\pi^-$ and opposite Collins asymmetries for oppositely charged hadrons are generated. Also, a dihadron asymmetry with the same sign as for positive hadrons is produced. StringSpinner uses the quantum mechanical formulation of this model, in which the spin effects depend on a complex parameter, tuned as in Ref. \cite{Kerbizi:2021pzn}. The initial quark polarisation is given by a parametrisation of the transversity PDF for valence $u$ and $d$ quarks. For this work we have used the default parametrisations, which were tuned to reproduce the $\pi^{+}$ and $\pi^{-}$ Collins asymmetries measured by COMPASS on an NH$_3$ target. The simulations were performed neglecting the intrinsic transverse momentum of the quarks, but it was checked that the dihadron asymmetries are not affected \cite{Kerbizi:2018qpp}.

\begin{figure}[tb]
\vspace{-1em}
\centering
\includegraphics[width=0.9\textwidth]{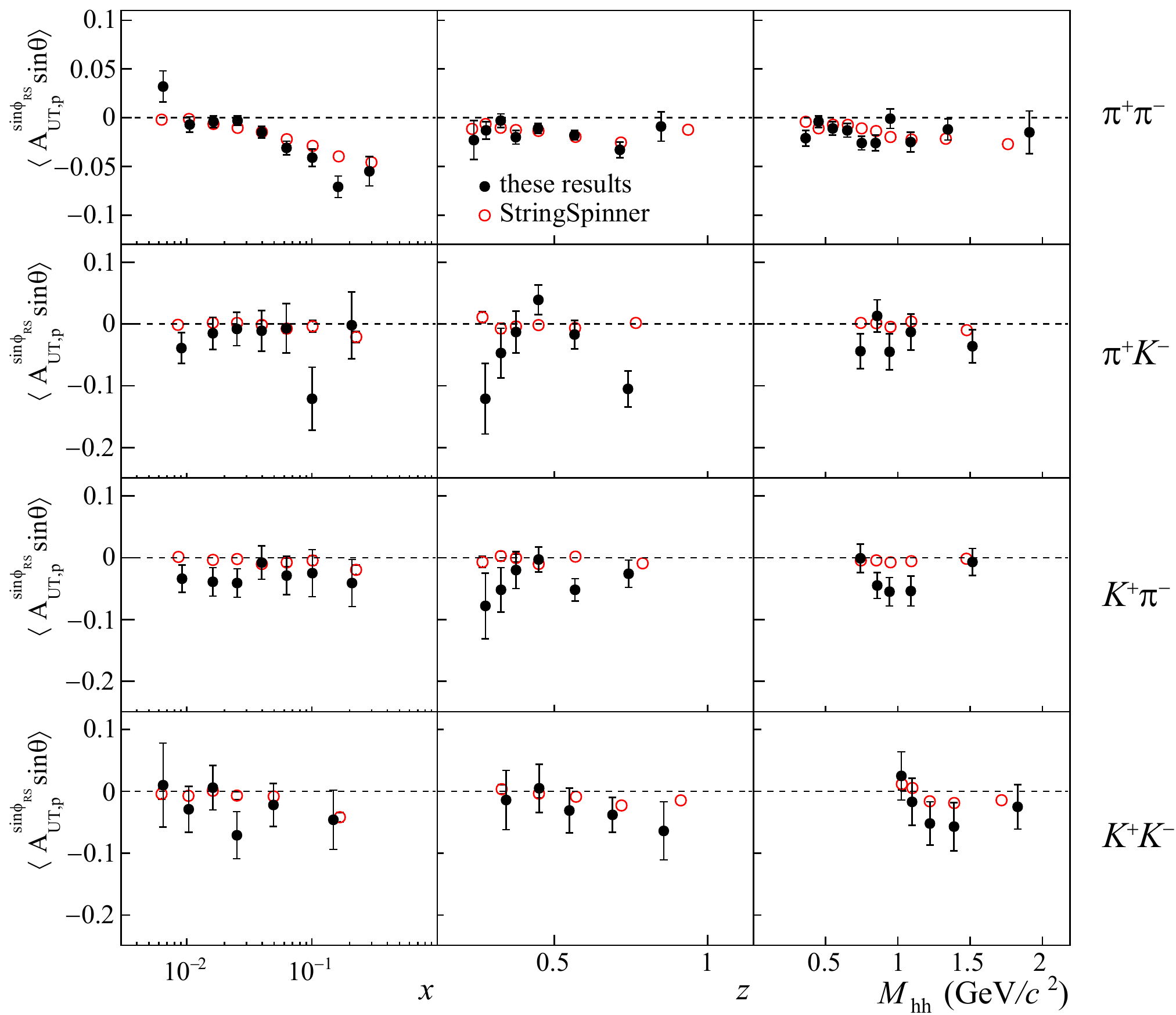}
\caption{Comparison between $\pi^+\pi^-$, $\pi^+K^-$, $K^+\pi^-$ and $K^+K^-$ asymmetries for proton data (closed points) and results from simulations using StringSpinner (open points).}\label{fig:StringSpinner}
\end{figure}

In Fig. \ref{fig:StringSpinner} the measured dihadron asymmetries (closed points) are compared to the simulated asymmetries (open points) for proton data. As can be seen, the simulation describes the data particularly well for $\pi^+\pi^-$ and $K^+K^-$ pairs, in all kinematic variables.
The trend of the asymmetries as a function of $x$ is mainly driven by the $x$-shape of the implemented transversity PDFs. While the $z$ and $M_{\rm hh}$ dependences are predictions of the model. The large signal for $\pi^+\pi^-$ and $K^+K^-$ pairs can be understood in the approximation of $u$-quark dominance considering the fact that $\pi^+$ or $K^+$ are most likely produced at rank one, whereas $\pi^-$ or $K^-$ are produced at rank two. Regarding the $\pi^+K^-$ and $K^+\pi^-$ pairs, the simulated asymmetries are small and compatible with the data within uncertainties. This is expected considering the fact that, \textit{e.g.}, the $\pi^+$ and the $K^-$ of a $\pi^+K^-$ pair are most likely produced at rank one and three separated by a rank two neutral kaon. Thus the $\pi^+$ and the $K^-$ are most likely emitted on the same side producing a small dihadron asymmetry.

In corresponding simulations for deuteron data, dihadron asymmetries compatible with zero were found for all types of hadron pairs. This is in agreement with the data and is expected from the fact that the transversity PDFs for valence $u$ and $d$-quarks have almost the same size but opposite sign.


\section{Conclusions}

In this paper we present the results of a new measurement of transverse-spin-dependent asymmetries in hadron pair production in DIS of 160 $\rm{GeV}/c$ muons off  transversely polarised deuteron ($^{6}$LiD) and proton (NH$_{3}$) targets. The measurement covers all possible combinations of oppositely charged pions and kaons observed in the COMPASS kinematic range. 

The deuteron data used in the analysis were collected during $2002$ and $2004$, while the proton data include two separate parts collected in 2007 and 2010. Both data sets were already used earlier to extract the Collins and Sivers asymmetries for semi-inclusively measured single hadrons, with separate publications for charged hadrons as well for identified pions and kaons. These two data sets are the largest ones available on this process, including \textit{e.g.} 28M (4M) pion pairs in the proton (deuteron) data, and they provide important input for global analyses. 

The proton data show significant non-zero asymmetries. For $\pi^+\pi^-$ pairs, values reach $-7\%$ in the region $x>0.032$ and $-2.5\%$ in the invarinat-mass region around the $\rho^0$-meson mass. Slightly negative asymmetries are observed for $K^+K^-$ and $K^+\pi^-$ pairs. The deuteron data show for all hadron combinations asymmetries compatible with zero, within statistical uncertainties.

\section*{Acknowledgements}
This work was made possible thanks to the financial support of our funding
agencies. We also acknowledge the support of the CERN management and staff, as
well as the skills and efforts of the technicians of the collaborating
institutes.

\newpage
\bibliographystyle{utcaps2}
\bibliography{hadron-pairs_ID}

\end{document}